\documentclass[aps, pra, twocolumn, amssymb, showpacs]{revtex4-1}
\usepackage{graphicx}
\usepackage{bm}
\usepackage[dvipsnames]{xcolor}
\definecolor{royalblue}{RGB}{40, 70, 200}
\usepackage[colorlinks=true, linkcolor=black, citecolor=black, urlcolor=royalblue]{hyperref}
\usepackage{float,stackengine}
\usepackage{mathtools}
\usepackage{adjustbox}
\usepackage{nicefrac} 
\usepackage{ifthen} 
\usepackage{amsthm} 

\newif\ifcolorsections

\usepackage{tikz}
\usetikzlibrary{calc,fadings,decorations.markings,decorations.pathreplacing,arrows,decorations.pathmorphing}
\tikzset{snake it/.style={decorate, decoration=snake}}
\tikzset{apply style/.code={\tikzset{#1}}}
\usepackage[pdftex,outline]{contour}
\newcommand{\halo}[3]{
\node[#1, transparency group, opacity=.7] at #2 {{\contour{white}{\color{white}{#3}}}};
\node[#1] at #2 {#3};
      }

\newcommand{\drawgrid}[4]{%
    \begin{scope}
        \draw[help lines, step=1, gray!50] (#1,#2) grid (#3,#4);
        \draw[help lines, step=5,black,thick] (#1,#2) grid (#3,#4);
        
        \draw[->] (#1,0) -- (#3+1,0) node[right,scale=5] {$x$};
        
        \draw[->] (0,#2) -- (0,#4+1) node[above,scale=5] {$y$};
        
        \pgfmathtruncatemacro{\xpfive}{#1 + 5};
        \foreach \x in {#1,\xpfive,..., #3}{
            \ifnum\x=0 \else \draw (\x,0) node[below,scale=5] {\x}; \fi
        }

        \pgfmathtruncatemacro{\ypfive}{#2 + 5};
        \foreach \y in {#2,\ypfive,...,#4} {
            \ifnum\y=0 \else \draw (0,\y) node[left,scale=5] {\y}; \fi
        }
    \end{scope}
}

\usepackage[normalem]{ulem}

\newcommand{\ket}[1]{\left|#1\right\rangle}
\newcommand{\bra}[1]{\left\langle#1\right|}

\def\cpm{%
  \mathchoice%
  {\xcpm\displaystyle{.2ex}{.53ex}}
  {\xcpm\textstyle{.2ex}{.53ex}}
  {\xcpm\scriptstyle{.16ex}{.43ex}}
  {\xcpm\scriptscriptstyle{.11ex}{.35ex}}
}

\def\xcpm#1#2#3{\mathbin{\ooalign{%
      \raise #2\hbox{\pdfliteral{1 0 0 rg}$#1+$\pdfliteral{0 g}}\cr
      \lower #3\hbox{\pdfliteral{0 0 1 rg}$#1-$\pdfliteral{0 g}}%
  }}}

\graphicspath{{Figures/}}
\usepackage{blindtext}
\usepackage{subfiles} 

\begin{document}
\title{Universal spectral bounds for the quantum Rabi model: Reformulating Braak’s conjecture}
\author{Alfonso Lanuza}
\thanks{lanuzagar@unizar.es}

\affiliation{Centro Universitario de la Defensa, Carretera de Huesca s/n, 50090 Zaragoza, Spain}

\date{\today}

\begin{abstract}
The quantum Rabi model is a paradigmatic example of a minimal yet nontrivial light-matter interaction, whose spectrum is transcendental yet exhibits a number of regularities. Braak observed that the eigenvalues bunch or anti-bunch following strict rules, leading to a conjecture that links integrability in quantum systems and residual order in their spectra. Understanding this structure is crucial for distinguishing deterministic quantum dynamics from chaotic behavior.  Here, we present violations of Braak's conjecture and reformulate it through a set of eigenvalue inequalities that conjecturally hold across all parameter regimes. We prove these bounds in the low splitting regime, we characterize the strongly-coupled widely-split limit, and provide universal upper bounds on the entire spectrum. Our results uncover additional layers of spectral organization in the quantum Rabi model and expand the analytic toolkit for strongly coupled quantum systems.
\end{abstract}

\maketitle

The quantum Rabi model (QRM) serves as a fundamental theoretical framework describing the dipolar interaction between a two-level system and a quantized bosonic mode \cite{Rabi1936, Rabi1937}. The deceptive simplicity of the QRM makes it a foundational building block across disciplines. It supplies a quantum description of light-matter coupling in cavity and waveguide quantum electrodynamics \cite{Haroche2006} and other atomic systems, such as atoms driven by a strong laser drive or nuclear spins in nuclear magnetic resonance, where the rotating-wave approximation fails giving rise to phenomena such as the Bloch-Siegert shift \cite{Bloch1940} and modifications of spontaneous emission rates \cite{Agarwal1971, Agarwal1973, Feranchuk2017}. The QRM finds implementations and analogues in ultracold atoms \cite{Koch2023}, cavity systems, superconducting circuits, trapped ions, quantum dots, organic molecules and hybrid quantum systems (see~\cite{Xie2017,Qin2024} and the references therein); and it is essential for understanding quantum phase transitions \cite{Hwang2015}, as counter-rotating terms not only can affect super- and subradiant decay rates \cite{Jorgensen2022} but also induce chaos in the quantum phase transition of the Dicke model \cite{Emary2003}. It has also become central to applications in quantum information science, particularly in nonclassical state generation and sensing in deep-strong coupling \cite{Kyaw2019}.\bigskip

Beyond its physical applications, the QRM also attracted attention towards its debated integrable structure \cite{Kus1985B, Braak2011, Moroz2013, Batchelor2015, Moroz2016, Xie2017} and the intricate nature of its spectral properties \cite{Feranchuk1996, Irish2007,Yu2012,Zhong2013,Maciejewski2014}. There are three disparate degrees of complexity associated with the QRM. Firstly, the Hamiltonian has a simple quadratic form 
\begin{equation}\label{eq:HR}
    H_R = \omega a^\dagger a + g \sigma_x (a + a^\dagger) + \Delta \sigma_z.
\end{equation}

We adopt the notation of \cite{Braak2011}, describing the two-level system by Pauli matrices $\sigma_{x,y,z}$ of level splitting $2\Delta$  coupled by $g$ to a quantum harmonic oscillator (QHO, presented in terms of the usual creation and annihilation operators $a^\dagger$ and $a$) of frequency $\omega>0$, and taking $\hbar=1$. Commutativity of the Hamiltonian with a parity operator \cite{Kus1985B, Feranchuk1996, Graham1984, Kus1985, Casanova2010} decouples its dynamics into two sectors $\mathcal{H}_\pm$, which can be visualized through the adjacency graph of the Hamiltonian \cite{Lanuza2024} (see Fig.~\ref{fig:QRM}).\bigskip

In their respective bases,
\begin{equation}
    \begin{array}{l}
         (\ket{\uparrow\! 0}, \ket{\downarrow\! 1}, \ket{\uparrow\! 2}, \ket{\downarrow\! 3},\ldots)\text{ for }\mathcal{H}_+\\
         \text{and }(\ket{\downarrow\! 0}, \ket{\uparrow\! 1}, \ket{\downarrow\! 2}, \ket{\uparrow\! 3},\ldots)\text{ for }\mathcal{H}_-,
    \end{array}
\end{equation}
these sectors (or parity chains \cite{Casanova2010}) have tridiagonal Hamiltonians
\begin{equation}\label{eq:Hpm}
    H_\pm = \begin{pmatrix}
    \pm\Delta & g\sqrt{1} & 0 & 0 &  \cdots \\
    g\sqrt{1} & \omega\mp\Delta & g\sqrt{2} & 0 &   \\
    0 & g\sqrt{2} &  2\omega\pm\Delta  & g\sqrt{3} &  \\
    0 & 0 & g\sqrt{3} & 3\omega\mp\Delta &  \\
    \vdots &   & &    & \ddots \\
\end{pmatrix}.
\end{equation}
The second level of complexity in the QRM consists of representing its spectral determinants, whose zeros form the spectrum of the system \cite{Dorey2007}. The spectral determinants can be expressed as continued fractions \cite{Schweber1967, Kus1986, Moroz2013, Braak2013}, three-term recurrences \cite{Feranchuk1996, Braak2011, Chen2011, Xie2017}, infinite products of $2\times2$ matrices \cite{Reik1982}, or in terms of confluent Heun functions \cite{Zhong2013, Maciejewski2014, Braak2019, Liang2025}. However, for numerical purposes, several authors \cite{Kus1985B, Nguyen2024, Xie2017} use as spectral determinants the characteristic polynomials of truncated versions of $H_\pm$; a method that has been demonstrated in the QRM despite the unboundedness of $H_\pm$  \cite{Braak2013}. The determinants of these truncated matrices can, in turn, be interpreted as components of the corresponding eigenvectors~\cite{Teschl1999}. In this way, finding the zeros of the spectral determinant becomes equivalent to allowing the eigenvector to be normalizable.\bigskip

Third, there is a considerable jump in complexity between the calculation of the spectral determinant and the spectrum. No general analytic expression for the eigenvalues is known, and their analytic properties continue to be the focus of ongoing investigation \cite{Charif2021}. When the boson and fermion are decoupled ($g=0$), the diagonal elements of $H_\pm$ in \eqref{eq:Hpm} form the spectrum, and the graph of the eigenvalues as functions of $\Delta$ is an infinite grid of two intersecting families of parallel lines of slope $\pm1$. Presumably, for $g\neq0$ sufficiently small, every level crossing becomes a conjugate pair of branch cuts in the complex $\Delta$ plane. This indicates that all energy levels $\{E^\pm_n(\Delta)\}_{n=0}^\infty$ are in fact different Riemann sheets of the same infinitely-valued function $E^\pm(\Delta)$. This analytic structure is non-elementary and transcendental, in the sense that no finite polynomial $c_n(\Delta)E^n+c_{n-1}(\Delta)E^{n-1}+\cdots+c_0(\Delta)$ can be written with entire coefficients in $\Delta\in\mathbb{C}$ whose zeros give values of the spectrum.\bigskip

\begin{figure}[t]
    \begin{center}
\resizebox{0.5\textwidth}{!}{\input{Tikzs/qRabiModel}}
\caption{Product states between a two-level system (on the top) and a quantum harmonic oscillator (on the right) are represented as red and blue dots, arranged according to their factors (height indicates the $\ket{n}$ factor, left for $\ket{\downarrow}$, and right for an $\ket{\uparrow}$ factor) and connected according to the transitions allowed by the Hamiltonian \cite{Lanuza2024}. The coupling $g$ between both subsystems generates two independent sectors $\mathcal{H}_\pm$.}
\label{fig:QRM}
	\end{center}
\end{figure}

As an alternative to the analytic approach, Ref.~\cite{Kus1985B} and more recently \cite{Nguyen2024} propose studying the eigenvalue spacings statistically, a technique leveraged to characterize quantum chaos in Hamiltonians \cite{Bohigas1984}, or pseudorandomness in number theory \cite{Montgomery1973} and in spectra of hyperbolic surfaces \cite{Sarnak2003}. Whereas signatures of chaos aren't present in the level spacing distribution of the QRM, this could be attributed to the lack of enough continuous degrees of freedom \cite{Nguyen2024}. Moreover, these signatures in the spectrum do appear when the number of two-level systems is increased \cite{Graham1986}, and  the semiclassical Rabi model is chaotic \cite{Milonni1983} even in the case of a single two-level system \cite{Gubernov2002}.\bigskip

The central question of this manuscript, then, is how to characterize a spectrum that is too intricate for an analytic description, yet too structured for a purely statistical one. Braak suggested that these hidden regularities might be encoded in inequalities rather than identities, a proposal now known as \textit{Braak’s conjecture} \cite{Braak2011} or $G$\textit{-conjecture} \cite{Nguyen2024}. For conciseness, if we denote by $B^\pm_n$ the number of eigenvalues $E^\pm$ of $H_\pm$ satisfying the inequalities \begin{equation}
    n\omega-g^2/\omega\leq E^\pm<(n+1)\omega-g^2/\omega,
\end{equation}
then Braak's conjecture states that $0\leq B^\pm_n\leq2$ and $1\leq B^\pm_n+B^\pm_{n+1}\leq3$ for $n=0,1,2,$... . This conjecture extends to arbitrary parameters the intimate connection between the spectrum of $H_\pm$ and that of a Lamb-shifted quantum harmonic oscillator ($E^\pm_n=n\omega-g^2/\omega$), which coincide when either $g\gg \Delta$ (deep strong coupling \cite{Casanova2010, Rossatto2017, Bustos2024}), $\Delta\ll\omega$ (weak level splitting or adiabatic limit \cite{Irish2007}), or in the high-energy limit $n\to\infty$, where the effect of $\Delta$ becomes negligible compared to the dominant energy scale. Using the asymptotic expansion of de Monvel and Zielinski \cite{Monvel2021}, Rudnick has recently established a density-one version of Braak's conjecture in the high-energy limit~\cite{Rudnick2024}. More precisely, he proved that the fraction of eigenvalues violating Braak's counting rules tends to zero as the energy increases.\bigskip 

There are, however, some limitations to Braak's conjecture. First, it refers to eigenvalue counts, but doesn't specify which eigenvalues fall in a given interval. Identifying the eigenvalues unambiguously is more effective and it becomes possible for $g\neq0$ since the tridiagonal structure of $H_\pm$ implies that energy levels do not cross \cite{Teschl1999oscillation}, which allows one to label them in strictly growing order $E^\pm_0<E^\pm_1<E^\pm_2<...$\ .\bigskip

Second, Braak's conjecture refers only to the positive part of the shifted spectrum, that lies between pole pairs of the $G$ function \cite{Braak2011}. In some parameter regimes there can be a large number of eigenvalues below the aforementioned region that are not covered by the conjecture. An example of this is the strongly-coupled widely-split limit $\lvert\Delta\rvert\!\sim\!\lvert g\rvert\gg \sqrt{n+1}\omega$ (see Fig.~\ref{fig:limit}  and App.~\ref{sec:limit}). In this limit, the quantization of the original QHO becomes negligible and Fock space acts as position space of a new QHO, where the spectrum is $E^\pm_n=n\omega-g^2/\omega- \Delta^2\omega/(4g^2)+O(\omega^2)$.\bigskip

\begin{figure}[t]
    \begin{center}
\resizebox{0.5\textwidth}{!}{\input{Tikzs/limit}}
\caption{a) Shifted eigenvalues $x_n^+$ (red) and $x_n^-$ (blue) of the Hamiltonian $H_\pm$ for $\Delta=4g$. The cyan dashed curve marks the position $\epsilon \coloneqq x/\omega+\Delta^2/(4g^2)={\left(4 g^4-\Delta ^2 \omega ^2\right)^2}/\left({64 g^6 \omega ^2}\right)$, corresponding to the point at which the lower edge of the Fock space intersects the effective harmonic potential [see panel (b)]. The cyan solid lines show $\epsilon=0,1,2,\,$... and indicate the quantized levels of the emergent QHO. b) Effective harmonic potential (black) for $\Delta=-4g=18\omega$. Colored curves show the eigenvector amplitudes $a_n$ (even $n$) and $b_n$ (odd $n$), shifted vertically by their corresponding effective energies $\epsilon$. Symbols denote numerical results obtained from a $600\times600$ truncation of the matrix representation in Eq.~\eqref{eq:Hpm}: filled circles correspond to eigenstates of $H_+$, while open circles correspond to eigenstates of $H_-$. Solid curves are the eigenstates of the effective QHO, normalized independently to match the analytic approximation of Eq.~\eqref{eq:eigenstate}. The QHO description breaks down when the Fock-space boundary at $n=1$ lies within the parabolic potential.}
\label{fig:limit}
	\end{center}
\end{figure}

Third, more relations beyond Braak's conjecture have been noted in the QRM, such as $1\leq B^+_n+B^-_n$ proposed in \cite{Nguyen2024}, suggesting that there is more order in the spectrum than originally anticipated.\bigskip

To overcome these limitations we propose the inequalities 
\begin{gather}
-\omega\leq E^+_n-E^-_n\leq \omega, \label{eBC:1}\\
E^\pm_{n+1}-E^\pm_n\leq 2\omega, \\
E^\pm_n\leq (n+1/2)\omega-g^2/\omega, \label{eBC:3}
\end{gather}
(for $n=0,1,2...$, see Fig.~\ref{fig:eigenvalues}) and strict counterparts if $g\neq0$.\bigskip

\begin{figure*}[t]
    \begin{center}
\resizebox{1.0\textwidth}{!}{\input{Tikzs/eigenvalues}}
\caption{Differences between eigenvalues $E^\pm_n$ of the QRM, obtained numerically through a $50\times50$ truncation of $H_\pm$ for variable coupling $g$ and several level splittings $\Delta$ (color-coded according to the scale shown at the lower right). Gray areas are conjectured to be off-bounds for the corresponding energy differences.}
\label{fig:eigenvalues}
	\end{center}
\end{figure*}

While Braak's conjecture also suggests the inequalities $E^\pm_{n+2}-E^\pm_{n}\geq \omega$ and $E^\pm_{n+3}-E^\pm_{n}\geq 2\omega$, violations can be found numerically by minimizing the level spacings $E^\pm_{n+k}-E^\pm_{n}$ for $k=2,3$. In turn, searching through violations of these inequalities leads to violations of Braak's conjecture, which we report in Table~\ref{tab:violations}.\bigskip

\begin{table}[t]
\centering
\begin{tabular}{cccc}
\noalign{\hrule height 1pt}
$g$ & $\Delta$ & Eigenvalues & Violation \\
\hline
6 & 30 & 
\begin{tabular}{@{}c@{}}
$E^{-}_{12}=-31.98930753$\\
$E^{-}_{13}=-31.26006761$\\
$E^{-}_{14}=-30.63548420$\\
$E^{-}_{15}=-30.05382413$
\end{tabular}
&
$B^-_4=B^-_5=2$
\\\hline
11.01 & \ 119\  & 
\begin{tabular}{@{}c@{}}
$E^{-}_{38}=-120.18343926$\\
$E^{-}_{39}=-119.65429442$\\
$E^{-}_{40}=-119.17307475$\\
$E^{-}_{41}=-118.60595603$\\
$E^{-}_{42}=-117.96683877$\\
$E^{-}_{43}=-117.28955999$
\end{tabular}
&
$B^-_1=B^-_2=B^-_3=2$
\\\hline
11.542 & \ 131.2\  & 
\begin{tabular}{@{}c@{}}
$E^{-}_{42}=-132.21769602$\\
$E^{-}_{43}=-131.71982373$\\
$E^{-}_{44}=-131.22727757$
\end{tabular}
&
$B^-_1=3$
\\
\noalign{\hrule height 1pt}
\end{tabular}
\caption{Parameter values (exact, with $\omega=1$) and eigenvalues (reported to eight decimal places) for which Braak's conjecture is violated.}
\label{tab:violations} 
\end{table} 

While proving bounds \eqref{eBC:1}-\eqref{eBC:3} on the spectrum exceeds the scope of this work, in the following we pinpoint analytical tools that help with this endeavor. Variational methods applied to physically-motivated ans\"atze \cite{Hwang2010,Ying2015,Leroux2017} are very effective in describing the ground state. A simpler version of the ones presented in the literature results in the upper bound
\begin{equation}\label{eq:variationalE0}
    \begin{aligned}
    &
E^+_0\leq
 \frac{g^{2}}{\omega\,\tanh\left(\tfrac{2g^{2}}{\omega^{2}}\right)}
 - \frac{2g^{2}\sqrt{1+\nu^{2}}}{\omega\sqrt{1-e^{-{4g^{2}/\omega^{2}}}}}
 \\&
 \text{where}\qquad
 \nu
=\frac
{ 1-\frac{\Delta\,\omega}{g^{2}} \sinh\!\left(\tfrac{2g^{2}}{\omega^{2}}\right) }{\sqrt{e^{4g^{2}/\omega^{2}}-1}},
    \end{aligned}
\end{equation}
that allows proving inequality~\eqref{eBC:3} for $n=0$ (see App.~\ref{sec:variationalBound} for details).\bigskip

On a different note, the displacement operator
\begin{equation}
D(\alpha)=e^{\alpha a^\dagger-\alpha^\dagger a}
\end{equation}
offers a unitary transformation
\begin{equation}
\begin{aligned}\label{eq:D(HR)}
    &D\left(\tfrac{g}{\omega}\sigma_x\right)H_R D\left(-\tfrac{g}{\omega}\sigma_x\right)=\omega a^\dagger a-\frac{g^2}{\omega}
    \\&
    +\tfrac{\Delta}{2}\,(\sigma_z + i \sigma_y)\, D\left(-\tfrac{2g}{\omega}\right)
+ \tfrac{\Delta}{2}\,(\sigma_z - i \sigma_y)\, D\left(\tfrac{2g}{\omega}\right)
\end{aligned}
\end{equation}
that diagonalizes the Hamiltonian when $\Delta=0$. Moreover, the operator accompanying $\Delta$ is similar to $\sigma_z$, and therefore has unit operator norm. By Weyl's theorem \cite{HornJohnson1985},
\begin{equation}
\left|E^\pm_n-n\omega+\frac{g^2}{\omega}\right|\leq\lvert\Delta\rvert
\end{equation}
implying \eqref{eBC:1}-\eqref{eBC:3} and $E^\pm_n\leq \min\left\{E^\pm_{n+2}-\omega,E_{n+3}^\pm-2\omega\right\}$ when $\lvert\Delta\rvert\leq\omega/2$. Consequently Braak's conjecture is correct in this regime.\bigskip

Gershgorin disks provide another powerful tool for analytically bounding an operator’s eigenvalues (see Ref.~\cite{HornJohnson1985} for the finite-dimensional case and Ref.~\cite{Aleksic2014} for the infinite-dimensional extension) that can be combined with other methods. In particular, the min-max theorem states that $E^\pm_n$ is bounded from above by the maximum eigenvalue of the truncation $(M_{jk})_{j,k=0}^n$ of \eqref{eq:D(HR)} restricted to $\mathcal{H}_\pm$, whose matrix elements are
\begin{equation}
M_{jk}=\left(j\omega-\frac{g^2}{\omega}\right)\delta_{jk}
    \pm(-1)^j\Delta \bra{k}D\left(\tfrac{2g}{\omega}\right)\ket{j}
\end{equation}
which can be rewritten in closed form using associated Laguerre polynomials \cite{Irish2005, Irish2022}
\begin{equation}
\begin{aligned}
   \bra{k}&D\left(\tfrac{2g}{\omega}\right)\ket{j}
=
e^{-2g^2/\omega^2}
\\&
\times\begin{cases}
\displaystyle
\sqrt{\frac{j!}{k!}}
\left(-\tfrac{2g}{\omega}\right)^{k-j}
L_j^{(k-j)}\!\left(\tfrac{4g^2}{\omega^2}\right)
&\text{if } k\geq j,\\[1.1em]
\displaystyle
\sqrt{\frac{k!}{j!}}
\left(\tfrac{2g}{\omega}\right)^{j-k}
L_k^{(j-k)}\!\left(\tfrac{4g^2}{\omega^2}\right)
&\text{if } j> k.
\end{cases}
\end{aligned}
\end{equation}
Then, by the Gershgorin circle theorem we obtain the bound
\begin{equation}
\begin{aligned}
  E^\pm_n\leq \max_{0\leq j\leq n}\Bigg\{
  j\omega &-\frac{g^2}{\omega}\pm (-1)^j\Delta 
\bra{j}D\left(\tfrac{2g}{\omega}\right)\ket{j}
  \\&
  +\lvert\Delta\rvert\sum_{\substack{k \neq j \\ k \le n}}
\left| \bra{k}D\left(\tfrac{2g}{\omega}\right)\ket{j}\right|
  \Bigg\}
  \\
  \qquad=n\omega-\frac{g^2}{\omega}\pm & \Delta e^{-\tfrac{2g^2}{\omega^2}}O\left(\frac{g^{2n}}{\omega^{2n}}\right)
\end{aligned}
\end{equation}
that converges super-exponentially to the deep-strong-coupling limit \cite{Feranchuk1996, Irish2005, Casanova2010,Rossatto2017}.\bigskip

In summary, we have introduced violations of Braak's conjecture and established a set of universal spectral inequalities that extend aspects of the conjecture by furnishing explicit, nonperturbative relations between eigenvalues across parity sectors and excitation numbers in the quantum Rabi model. These bounds are rigorously demonstrated, together with Braak's conjecture, in the low‑splitting regime $\lvert \Delta\rvert\leq\omega/2$ via displacement techniques and Weyl‑type perturbative arguments. A variational approach is employed to establish the proposed upper bound for the ground-state energy, while Gershgorin disks provide upper bounds for all energy levels, although these bounds are primarily useful in the deep-strong-coupling regime $g/\omega\gg1$. In addition, the harmonic‑oscillator approximation developed for the strongly-coupled widely-split regime clarifies eigenvector localization and energy shifts, yielding a coherent picture across parameter extremes. Taken together, these results reveal previously unrecognized spectral organization in a model of intrinsic transcendental complexity and offer analytic tools that may inform broader studies of nonperturbative, integrable, and near‑integrable quantum systems.

\section{Acknowledgments}
I wish to thank J.  de Ram\'on and A. Gonz\'alez-Tudela for insightful discussions and A. Ballesteros for carefully reading the manuscript and providing valuable feedback. I am grateful to the \textit{Quantum} referees for their thoughtful and constructive reports. In particular, Referee 2 discovered Braak's violation at $\Delta=5g=30\omega$, leading the author to find the other violations presented in Table~\ref{tab:violations}. This work has been supported by the Q-CAYLE Project funded by the Regional Government of Castilla y Le\'on (Junta de Castilla y Le\'on) and the Ministry of Science and Innovation MICIN through NextGenerationEU (PRTR C17.I1), as well as from the grant PID2023-148373NB-I00 funded by MCIN/AEI/ 10.13039/501100011033/FEDER -- UE.

\appendix

\section{Strongly-coupled widely-split limit}
\label{sec:limit}
In recent years, the strong and ultrastrong coupling regimes of the QRM \cite{Qin2024} have attracted significant interest due to their experimental realizations in superconducting circuits, cavity and circuit quantum electrodynamics, and solid-state quantum systems \cite{Niemczyk2010,Yoshihara2017,Forn-Diaz2017}. Notably, there is no single ``strong coupling limit''; rather, multiple regimes emerge depending on the relation between the coupling strength, the oscillator frequency, and the spectral region under investigation \cite{Pedernales2015,Rossatto2017}. To the best of our knowledge, a complete characterization of these coupling regimes remains elusive, particularly when the two-level splitting is comparable to the coupling strength. In this section we characterize the lower part of the spectrum and its corresponding eigenvectors in this strongly-coupled widely-split limit.\bigskip

For some energy $E$, let us denote an eigenvector of $H^+$ in its matrix representation \eqref{eq:Hpm} by coordinates $a_n$ if $n$ is even and $b_n$ if $n$ is odd. Then by the eigenvalue equation,
\begin{equation}\label{eq:absystem}
    \left\{\begin{array}{c}
         g\sqrt{n}b_{n-1}+\left(n\omega+\Delta-E\right)a_{n}+g\sqrt{n+1}b_{n+1}=0  \\
          g\sqrt{n}a_{n-1}+\left(n\omega-\Delta-E\right)b_{n}+g\sqrt{n+1}a_{n+1}=0.
    \end{array}
    \right.
\end{equation}
After a simple substitution,
\begin{widetext}
\begin{equation}\label{eq:anbn}
    \left\{\begin{array}{c}
         \frac{g^2 \sqrt{n-1} \sqrt{n}}{E + \Delta + \omega - n \omega} a_{n-2} 
+ \left( -E + \Delta + n \omega + \frac{g^2 n}{E + \Delta + \omega - n \omega} 
+ \frac{g^2 (n+1)}{E + \Delta - (n+1) \omega} \right) a_n 
+ \frac{g^2 \sqrt{n+1} \sqrt{n+2}}{E + \Delta - (n+1) \omega} a_{n+2} = 0
\\
          \frac{g^2 \sqrt{n-1} \sqrt{n}}{E - \Delta + \omega - n \omega} b_{n-2} 
+ \left( -E - \Delta + n \omega + \frac{g^2 n}{E - \Delta + \omega - n \omega} 
+ \frac{g^2 (n+1)}{E - \Delta -(n+1) \omega} \right) b_n 
+ \frac{g^2 \sqrt{n+1} \sqrt{n+2}}{E - \Delta - (n+1) \omega} b_{n+2} = 0.
    \end{array}
    \right.
\end{equation}
\end{widetext}
To solve for $a_n$ (the case for $b_n$ is analogous) in the limit $\omega\ll \lvert g\rvert, \lvert \Delta\rvert$, we can make the change to a continuous variable
\begin{equation}
\begin{aligned}
    n&=\frac{\nu}{\omega }-\frac{\Delta ^2}{4 g^2}+\frac{g^2}{\omega ^2}
    \qquad
    E=\epsilon\omega-\frac{\Delta ^2 \omega }{4 g^2}-\frac{g^2}{\omega }
    \\
    a_n &=a(\nu)
    \\
    a_{n\pm 2} &=a(\nu\pm 2\omega)=a(\nu)\!\pm\! a'(\nu) 2\omega\!+\!a''(\nu)2\omega^2\!+\!O\!\left(\omega^3\right),
\end{aligned}
\end{equation}
and make a Taylor expansion of equation \eqref{eq:anbn} in $\omega$, finding that the orders at $\omega^{-1}$ and $\omega^0$ cancel for consistency, whereas the next leading order cancels when
\begin{equation}
    \left(\frac{\nu^2}{2 g^2} -2 \epsilon-1\right)a(\nu)-2 g^2 a''(\nu)=0.
\end{equation}
This differential equation is easily recognizable as the eigenvalue equation of a quantum harmonic oscillator, which has square-integrable solutions only when $\epsilon=0,1,2,$... and
\begin{equation}
a(\nu) \propto \sqrt{\frac{\lvert\omega/g\rvert}{2^\epsilon \epsilon!}} \left({2\pi} \right)^{-1/4} 
e^{-\frac{\nu^2}{4 g^2}} \mathrm{He}_\epsilon \left(\frac{\nu}{\sqrt{2}g} \right),
\end{equation}
where $\mathrm{He}_\epsilon$ denotes a Hermite polynomial.\bigskip

We note that for $\epsilon\gtrsim {\left(4 g^4-\Delta ^2 \omega ^2\right)^2}/\left({64 g^6 \omega ^2}\right)$ the approximation is compromised because the function starts to have a significant overlap with values of $\nu$ that correspond with negative values of $n$; but this upper region of the spectrum is avoided if $\lvert\Delta\rvert\!\sim\!\lvert g\rvert\gg\sqrt{\epsilon+1}\omega$. The expressions for $b_n$ (or $b(\nu)$) are identical since these results don't depend on the sign of $\Delta$; however, there is a different constant of proportionality for $b(\nu)$. This mismatch $a_{n_0}/b_{n_0+1}$ can be estimated now that we know that the state is localized around $n_0=g^2/\omega^2-\Delta^2/(4 g^2)$ with the rough approximation $\sqrt{n_0}b_{n_0-1}\approx\sqrt{n_0+1}b_{n_0+1}$ in the first equation of \eqref{eq:absystem}, resulting in
\begin{equation}
\begin{aligned}
    \frac{a_{n_0}}{b_{n_0+1}}&\approx-\frac{2g\sqrt{n_0}}{n_0\omega+\Delta-E}
    \\
    &=\text{sign}(g) \left(-1+\frac{\Delta  \omega }{2 g^2}+O\!\left(\omega ^2\right)\right).
\end{aligned}
\end{equation}
Altogether (see Fig.~\ref{fig:limit}) we have that, in the strongly-coupled widely-split limit,
\begingroup
\thinmuskip=0mu \medmuskip=0mu \thickmuskip=0mu
\begin{equation}\label{eq:eigenstate}
\begin{aligned}
    \Bigg(&\left[\tfrac{1-(-1)^n}{2}\left(1\pm\tfrac{\Delta  \omega }{4 g^2}\right)+\tfrac{1+(-1)^n}{2}\text{sign}(g)\left(-1\pm\tfrac{\Delta  \omega }{4 g^2}\right) \right] 
    \\&\times
    \sqrt{\tfrac{\lvert\omega/g\rvert}{2^\epsilon \epsilon!}} (-1)^\epsilon\left({2\pi} \right)^{-1/4} 
e^{-\left( \tfrac{n\omega}{2g} + \tfrac{\Delta^2\omega}{8g^3} - \tfrac{g}{2\omega}\right)^2}
\\&\times
\mathrm{He}_\epsilon \left(\tfrac{n\omega}{\sqrt{2}g} + \tfrac{\Delta^2\omega}{4\sqrt{2} g^3} - \tfrac{g}{\sqrt{2}\omega}\right)\Bigg)_{n=0}^\infty
\end{aligned}
\end{equation}
\endgroup
approximates the eigenvector of $H_\pm$ of energy $E_\epsilon=\epsilon\omega-\frac{g^2}{\omega }-\frac{\Delta^2 \omega }{4 g^2}+O(\omega^{2})$ where $\epsilon$ can take non-negative integer values.

\section{A variational bound for the ground energy}
\label{sec:variationalBound}
A variational ansatz must include enough structure to represent the relevant features of the ground state, while remaining sufficiently simple to allow analytical manipulation. Although variational ans\"atze for the QRM ground state have been studied in detail \cite{Hwang2010,Ying2015,Leroux2017}, achieving the target bound $E_0^\pm\leq\omega/2-g^2/\omega$ does not require the accuracy or number of parameters used in previous works. Our variational ansatz for the ground state of $H_+$ is
\begin{equation}
    \left(\mu_{(-1)^n}\frac{(-g/\omega)^n}{\sqrt{n!}}e^{-\frac{g^2}{2\omega^2}}\right)_{n=0}^\infty
\end{equation}
that depends on the free parameters $\mu_+$ and $\mu_-$, restricted by the normalization condition
\begin{equation}
    \mu_+^2 e^{-\frac{g^2}{\omega^2}}\cosh\left(\frac{g^2}{\omega^2}\right)+\mu_-^2 e^{-\frac{g^2}{\omega^2}}\sinh\left(\frac{g^2}{\omega^2}\right)=1.
\end{equation}
The parameter values that minimize the energy of this state are
\begin{equation}
\mu_\pm=\sqrt{\frac{1\pm\frac{\operatorname{sign}\nu}{\sqrt{1+\nu^{-2}}}}{1\pm e^{-2g^2/\omega^2}}}
\end{equation}
where $\nu$ and the minimal energy are presented in \eqref{eq:variationalE0}. To prove that this energy is lower than $\omega/2-g^2/\omega$, we note that
\begin{equation}
 \sqrt{1+\nu^{2}}\geq 1
\end{equation}
and thus
\begin{equation}
    \begin{aligned}
E^+_0&\leq
 \frac{g^{2}}{\omega\,\tanh\left(\tfrac{2g^{2}}{\omega^{2}}\right)}
 - \frac{2g^{2}}{\omega\sqrt{1-e^{-\frac{4g^{2}}{\omega^{2}}}}}
 \\& \leq
 \frac{g^2}{\omega}+\frac{\omega}{2}
 - \frac{2g^{2}}{\omega\sqrt{1-e^{-\frac{4g^{2}}{\omega^{2}}}}}\leq \frac{\omega}{2}-\frac{g^2}{\omega}
 \end{aligned}
\end{equation}
 which extends to $E_0^-$ by the symmetry between both sectors when $\Delta\leftrightarrow-\Delta$.

\bibliographystyle{apsrev4-1}
\bibliography{QRM}

\end{document}